\documentclass{article}
\usepackage{ijcai24}

\usepackage{times}
\usepackage{soul}
\usepackage{url}
\usepackage[hidelinks]{hyperref}
\usepackage[utf8]{inputenc}
\usepackage[small]{caption}
\usepackage{graphicx}
\usepackage{amsmath}
\usepackage{amsthm}
\usepackage{amssymb}
\usepackage{booktabs}
\usepackage{algorithm}
\usepackage{algorithmic}
\usepackage[switch]{lineno}
\usepackage[frozencache=true,cachedir=minted-cache]{minted}
\usepackage{natbib}
\usepackage[dvipsnames]{xcolor}

\usepackage{todonotes}
\usepackage{multirow}

\title{Out of style: Misadventures with LLMs and code style transfer}

\author{
Karl Munson$^1$
\and
Chih-Kai Ting$^1$\and
Serenity Wade$^1$\and
Anish Savla$^1$\and
Julian Dolby$^2$\and
Kiran Kate$^2$\And
Kavitha Srinivas$^2$
\affiliations
$^1$University of California, Santa Cruz, USA\\
$^2$IBM Research, USA\\
\emails
\{ksmunson, cting3, swade1, ansavla\}@ucsc.edu, dolby@us.ibm.com, kavitha.srinivas@ibm.com, kakate@us.ibm.com
}

\begin{document}

\maketitle

\begin{abstract}
Like text, programs have styles, and certain programming styles are more desirable than others for program readability, maintainability, and performance.  Code style transfer, however, is difficult to automate except for trivial style guidelines such as limits on line length.  Inspired by the success of using language models for text style transfer, we investigate if code language models can perform code style transfer.  Code style transfer, unlike text transfer, has rigorous requirements: the system needs to identify lines of code to change, change them correctly, and leave the rest of the program untouched.  We designed CSB (Code Style Benchmark), a benchmark suite of code style transfer tasks across five categories including converting for-loops to list comprehensions, eliminating duplication in code, adding decorators to methods, etc. We then used these tests to see if large pre-trained code language models or fine-tuned models perform style transfer correctly, based on rigorous metrics to test that the transfer did occur, and the code still passes functional tests. Surprisingly, language models failed to perform all of the tasks, suggesting that they perform poorly on tasks that require code understanding.  We will make available large-scale corpora to help the community build better code models.
\end{abstract}

\section{Introduction}
As with text, programs have style attributes, and certain styles are more desirable than others for program readability, maintainability, and even performance (\cite{naturalize}).  Most enterprises consider program style so important that they mandate adherence to organization-specific styles; in fact, \cite{naturalize} point out that one-third of manual code reviews are about adherence to style.  The process of reviewing and transferring across program styles, however, is a difficult problem to automate using symbolic approaches.  As a result, enterprises rely mostly on manual code reviews to enforce style requirements; only the most syntactic style constraints such as new lines after method calls or lines that do not exceed a certain length can be checked and performed automatically.

Inspired by work on text style transfer, we ask whether it is possible to use neural models for program style transfer.  Existing work on text style transfer shows that neural models perform well on text transfer (\cite{riley2021textsettr}, \cite{Tikhonov_2019}, \cite{zero_shot_text_style_transfer}, \cite{tst_survey}, \cite{reif-etal-2022-recipe}), where a single sentence is translated into a different styled sentence.

\begin{figure}
	\centering
	\includegraphics[width=.5 \textwidth]{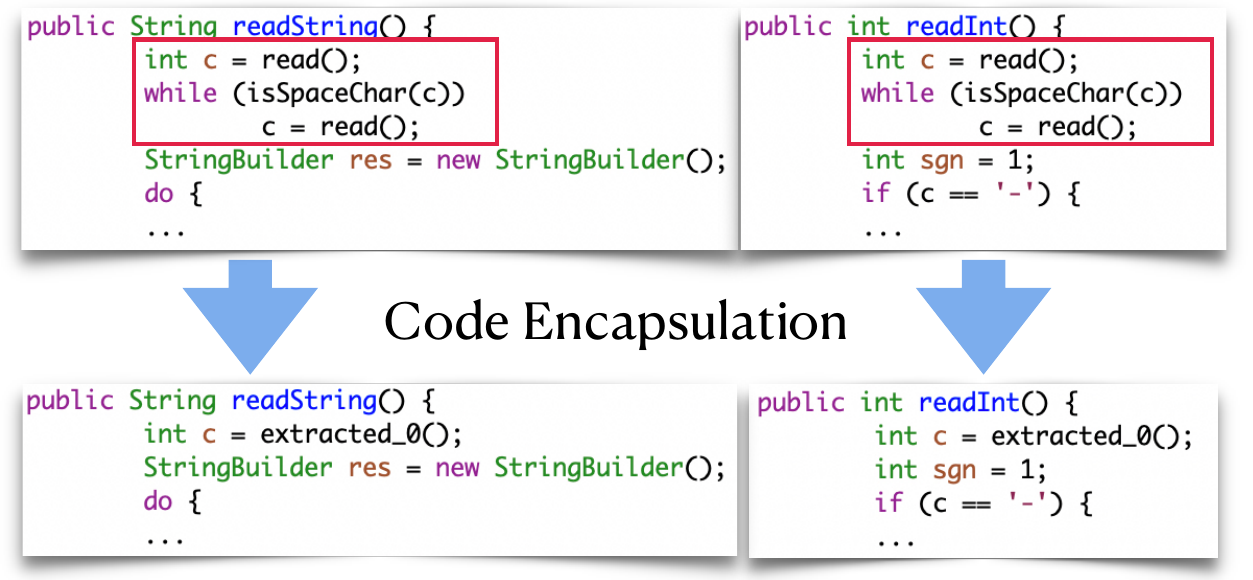}
	\caption{An example of the code encapsulation task.}
	\label{fig:style_example_fig}
\end{figure}

However, program style transfer has different challenges compared to text style transfer:
\begin{itemize}
    \item Unlike existing research on text transfer, which tends to require that a short single sentence be transformed as a whole into another sentence, style transfer in programs involves identifying one or more specific snippets of code in a much larger script to replace with an alternative while leaving large parts of the code unchanged. \textit{Most importantly, the semantics of programs are formally defined, which means that the tasks can serve as a stress test to the capabilities of language models.}
    \item Evaluation of text style transfer is subjective because precisely checking meaning preservation, as well as style transfer strength, is difficult for text. For code, there are rigorous metrics one can employ to measure how well programs perform style transfer.
\end{itemize}


Figure~\ref{fig:style_example_fig} illustrates an example of these challenges in a code encapsulation transfer task. The methods \emph{readString} and \emph{readInt} in the example have the same first 3 lines of code outlined in red. These lines are duplicated and encapsulating them into a new method \emph{extracted\_0} is better programming practice. To do so, however, requires a non-trivial level of code understanding; duplicate lines need to be identified and moved into a new method, and the method needs to be called correctly with the right parameters. Additionally, all other lines need to be left as is.

The goal in this work was to examine in detail, the capabilities of large language models for code.  Although language models have been shown to perform well on code generation tasks, we found that existing language models do not perform code style transfer at all.  Surprisingly, even seemingly simple style transfer tasks such as casing were difficult for language models.  Casing of variables seems trivial, but it does require an understanding of program scope, potential naming conflicts with existing variables, and appropriate changes in references to renamed variables.  In addition, even when models changed the required lines correctly, they often changed other lines of code they should have left untouched.  It was also fairly common for language models to 'pass' functional tests, by not transforming the code at all.  We therefore designed a new metric DiffCorrect that combines functional correctness with diffs to provide insight into these types of errors, to help us understand exactly what language models do with code.  

To document and understand code style effects more carefully, we designed a set of benchmarks for code called the Code Style Benchmark (CSB) in two different languages (Python and Java).  Each program had one of five classes of style changes performed on it, ensuring that changing the program's style did not change the semantics of the program. We generated parallel corpora for (a) changing casing conventions for a language (e.g. changing variables to lowercase), (b) creating list comprehensions from for-loops for Python, (c) eliminating duplicate code in a program for reuse, (d) adding documentation to code, and (e) adding decorators to methods.  We note that these are useful style transfer tasks that would be nice to have in IDEs but in fact, no IDEs exist yet to do this because many transformations such as even casing are hard to perform algorithmically.\footnote{Refactoring to extract code into methods does exist in IDEs, but, since it requires explicit code ranges to extract, it is intractable to apply it exhaustively to eliminate duplication.} We also note that to generate the parallel corpora, we can easily undo casing or eliminate documentation, but the actual desired stylistic convention (i.e., to have correctly cased variables in code) is difficult to generate.  To our knowledge, our benchmark is the first attempt at defining a dataset for code transformations that are potentially useful features that can be added to IDEs; the purpose of this dataset is to provide a rigorous benchmark to evaluate if language models can perform such transformations correctly.  We evaluated all semantics preserving transforms of the code on the CodeNet (\cite{codenet}) benchmark because that benchmark has functional tests.  

 The contributions of this paper can be summarized as follows:
 \begin{itemize}
   \item[(i)] We tested whether large language models understand code well enough to perform code style transforms while preserving the semantics of code. 
  \item[(ii)] We defined a code style benchmark (CSB) across 2 programming languages. CSB has parallel corpora for 5 style transform tasks and the dataset creation has been rigorously tested for functional correctness on the CodeNet dataset. To the best of our knowledge, CSB is the first benchmark for code style transfer. We will release the datasets publicly for further research on code style transfer.
  \item[(iii)] Based on the observation that functional tests can be passed while not performing any transform at all, we define a new metric called DiffCorrect that is broadly applicable to code generation tasks where the output is a slightly modified version of the input.
  \item[(iv)]  The code to perform the transforms and generate parallel corpora is provided as a toolkit, so new datasets can be generated for the purpose of training LLMs for the future.  To help increase the diversity of programs for training, we also ran transforms on a large Python training dataset and used it for fine-tuning.  The training datasets are available at \href{https://zenodo.org/record/7829238}{here}, and the code is available \href{https://anonymous.4open.science/r/codeStyle-0C41}{here}.
\end{itemize}
 

\section{Related Work / Background}
\subsection{Code Style}
Style is recognized as an important aspect of programming (\cite{style_def}), and coding style guides usually have a list of recommendations. IDEs and code formatting tools enable automatic style transformations for line lengths, spacing, etc. In this paper, we focus on complex aspects of style based on language features commonly used. Specifically, our work was informed by the analysis of common language features in popular open-source projects, performed by \cite{9425916} and \cite{9796194}. Their work, however, does not address the style transfer task. \cite{naturalize} uses statistical techniques to identify styles from code bases for Java code and suggest single identifier names and formatting. The style transformations in CSB are less about identifier names and more about a set of language features that define style for a programming language; casing might be thought to be similar to the work in \cite{naturalize}, but the goal of \cite{naturalize} is to predict entire variable names rather than change their case. \cite{Mariano_oopsla22} propose a new neural network architecture to convert from imperative code snippets to their equivalent functional code. This task is similar to the list comprehension task we include in CSB, but the authors do not focus on a general style transfer benchmark.

\subsection{Code Language Models and Tasks}
Self-supervised training of language models has been extended to code and has led to a variety of pre-trained models. There are encoder-only pre-trained models such as 
CodeBERT \cite{feng2020codebert}, CuBERT \cite{cubert}, and C-BERT \cite{DBLP:journals/corr/abs-2006-12641}, encoder-decoder models such as CodeT5 \cite{codet5} and PLBART \cite{plbart} and generative models such as Codex \cite{codex}, Starcoder \cite{li2023starcoder}. We use CodeT5 for fine-tuning to demonstrate the usefulness of our benchmarks in terms of fine-tuning and Starcoder to see if very powerful neural models can be used with prompting to transform program style.

These models have been successfully applied to many code tasks such as 
bug detection, clone detection, translation, code search, documentation generation, code completion, code summarization, variable misuse, incorrect use of operators or operands, etc. There are benchmark datasets available for these tasks including CodexGLUE \cite{DBLP:journals/corr/abs-2102-04664}, and CodeSearchNet \cite{https://doi.org/10.48550/arxiv.1909.09436}. To the best of our knowledge, style transfer for list comprehension, decorators, casing, and code encapsulation are novel tasks and there are no parallel corpora available for them. Documentation generation and its related task of comment generation are well known but we include it in our style transfer benchmark because docstrings and comments are often considered desirable style attributes of code.

\section{Benchmark Task Definitions}
\label{tasks}
This section explains the tasks under CSB and provides details of how we created the parallel corpora for each task.


 In most cases, it can be observed it is easier to transform the code in one direction, but not the other.  As an example, it is easy to convert an existing list comprehension to a for-loop; the reverse is harder because not every for-loop can be written as a list comprehension. For this case, our target style is functional i.e. with list comprehensions, but we do the reverse transformation to obtain a parallel corpus for that task. For most of the tasks we considered, we modified the original code, and then used the transformed code as input for the models, with the original being the gold label.  The one exception was refactoring duplicate code in Java, where we exploited existing tools to refactor the code in the correct direction.  We nevertheless include it in the benchmark because, although such refactoring tools exist in Java, it requires the programmer to select the code to be refactored.  We exhaustively searched lines of code to compute duplication; this is possible only for smaller programs.  Our goal here is to have neural models learn it for small programs, but then apply it to larger ones, and better yet generalize the transform to dynamically typed languages where such refactoring is even more difficult.  To test the correctness of the transforms, we tested against the Python and Java subsets of the CodeNet (\cite{codenet}) corpus, so each transform was checked for correctness by verifying that the output of the transformed program on the specified inputs matches the output of the original program.  

\subsection{List Comprehensions}
List comprehension is a popular functional programming feature in Python, and is often considered a representative of idiomatic Python (\cite{pythonic}). List comprehensions make the code more readable and also execute efficiently (\cite{9796194}). It is an example of a transform that is relatively straightforward to convert the comprehension to a for-loop, but not all for-loops can be converted into comprehensions, and that is why a statistical approach to the style change is useful. Figure~\ref{list_comp_example} shows an example of such a transform on a Python snippet; the list comprehension on line 11 gets converted into a for loop in lines 11-13. To transform list comprehensions, we traverse the AST iterating through the body of nodes looking for instances of ast.Assign and where the value of the body is ast.ListComp. An equivalent for-loop to the list comprehension is created.

\begin{figure}
\begin{minted}{python}
import fractions
from functools import reduce

def lcm_base(x, y):
    return (x * y) // fractions.gcd(x, y)
    
def lcm_list(numbers):
    return reduce(lcm_base, numbers, 1)
 
n=int(input())
t=[int(input()) for i in range(n)] 
 
print(lcm_list(t))
\end{minted}
\hfill
\begin{minted}{python}
import fractions
from functools import reduce

def lcm_base(x, y):
    return (x * y) // fractions.gcd(x, y)

def lcm_list(numbers):
    return reduce(lcm_base, numbers, 1)
    
n = int(input())
t = []
for i in range(n):
    t.append(int(input()))

print(lcm_list(t))
\end{minted}
\caption{Original program and the list comprehension transform}
\label{list_comp_example}
\end{figure}




\subsection{Decorators}  
Using decorators to add functionality to existing functions is a commonly used feature in Python (\cite{9425916}). To transform code to eliminate decorators, we eliminated each AST node's decorator list in the Python AST and rewrote the AST tree back to the source. Decorators were removed from ast.FunctionDef, ast.ClassDef, and ast.AsyncFunctionDef nodes and this was applied recursively to child nodes.   Figure~\ref{decorator_example} shows an example of the original program with a decorator on line 4 that is eliminated by the transform. Note here that the code does contain imports on line 2 which should help the model reconstruct the missing decorator.  This is typically the case with Python.  Decorators in Python do sometimes change the semantics (e.g., \texttt{classmethod}), while others like \texttt{lru\_cache} are hints for the performance characteristics of the program.  We therefore kept only the decorators that do not change the semantics of code, which were \texttt{lru\_cache} and \texttt{njit}. 

We performed the analogous transformation in Java, eliminating annotations.  To preserve program meaning, we eliminated only those annotations that are defined to be \texttt{@Retention(value=SOURCE)}; in the CodeNet programs, that comprised \texttt{@Override} and \texttt{SupressWarnings}.


\begin{figure}
\begin{minted}{python}
def d_xor_sum(N):
    from functools import lru_cache

    @lru_cache()
    def f(k):
        return (1, 2)[k] if k <= 1
        else f(k // 2) + f((k - 1) // 2)
        + f((k - 2) // 2)

    ans = f(N) % (10**9 + 7)
    return ans

N = int(input())
print(d_xor_sum(N))
\end{minted}
\hfill
\begin{minted}{python}
def d_xor_sum(N):
    from functools import lru_cache

    def f(k):
        return (1, 2)[k] if k <= 1
        else f(k // 2) + f((k - 1) // 2)
        + f((k - 2) // 2)
    ans = f(N) % (10 ** 9 + 7)
    return ans
    
N = int(input())
print(d_xor_sum(N))
\end{minted}
\caption{Original program and its decorator transform}
\label{decorator_example}
\end{figure}

\subsection{Casing} 
Consistent casing of identifier names is an important programming convention that improves readability. 
We created two different versions of this task for Java and Python, by focusing on the casing of variables alone.  To create parallel corpora, we identified assignments to variables using the AST parser for Python and the Eclipse refactoring for Java; in both cases, we lower-cased them and stripped out underscores.  In identifying assignments that should be changed, we ensured that we did not rename the variable to one that already existed in the program scope, rename it to a reserved keyword in the language, alias to method names, or alias to class names within the program.  All references to that variable within the program scope were then renamed to the same variable.  Figure~\ref{casing_example} illustrates how case is changed carefully, to preserve program semantics.  \texttt{MOD} is left unchanged because it is referenced as a parameter, all other variables are changed, consistently within scope.  

For Java, we used Eclipse refactoring tools to perform renaming.  Eclipse gives us two things: a way to iterate across variable names only, and it ensures that renamed variables do not cause conflicts with existing variables, do not become a keyword, etc.  Thus our renaming just iterates across all variables in each method, trying to convert it to lowercase and skipping variables for which the refactoring fails.  

\begin{figure}[ht]
\centering
\begin{minted}{python}
S = np.frombuffer(read().rstrip(),'S1')

MOD = 10**9 + 7

Sright = (S==b'>')
Sleft = (S==b'<')
Sadd = (S==b'+')
Ssub = (S==b'-')

def cumprod(arr,MOD):
    L = len(arr)
\end{minted}
    \hfill
\begin{minted}{python}
s = np.frombuffer(read().rstrip(), 'S1')

MOD = 10 ** 9 + 7

sright = (s == b'>')
sleft = (s == b'<')
sadd = (s == b'+')
ssub = (s == b'-')

def cumprod(arr, MOD):
    l = len(arr)
\end{minted}
    \caption{Original program, and its casing transform}
\label{casing_example}
\end{figure}

\subsection{Documentation} 
 Documentation for modules, classes, functions, and methods is commonly specified as docstrings in Python. Inferring documentation from code is a well-known task \cite{codexglue}, \cite{codex}, and we include it here because it is an important attribute of coding style. Dataset generation was performed via the removal of docstrings using a Python AST parser. This was implemented by traversing the AST for ast.Module, ast.FunctionDef, ast.ClassDef, and ast.AsyncFunctionDef nodes. For each of these nodes, docstrings were removed by checking if the first part of the body of the nodes listed above is an expression with a string value. We do not include actual examples here because the deletion is obvious.  
 
\subsection{Code Reuse} 

Good style is often taken to imply little redundancy, so we created a task of removing redundancy.  Since code equivalence in general is undecidable, we focus on a specific task: extracting duplicate chunks of code into a method. For constructing our dataset, this refactoring must be automatable across a repository of code.  To our knowledge, the combination of non-trivial method extraction and automation is best supported by Eclipse JDT, so this task applies only to our Java code.

In particular, we apply the Extract Method refactoring to every contiguous range of code of more than one statement, and Eclipse determines whether it can be extracted into a method.  If so, Eclipse also finds any other chunks of code that can use the same function.  If there is at least one such chunk, we execute the refactoring, yielding a new method with at least two calls.

This approach requires an exhaustive search across possible code chunks, and, for each of them, a search for duplicates.  As such, it is very expensive and suitable only for tasks such as dataset curation, where running for hours is acceptable.  Even then, the search is, at best, quadratic in program size, so it cannot scale to large programs. It is sufficient for CodeNet Java. 

\begin{figure}[h]
\centering
\begin{minted}{java}
public class Main {
	static public void main(String args[]){
		Scanner sc=new Scanner(System.in);
		 
		 int n=sc.nextInt();
		 int m=sc.nextInt();
		 if(m<2)m=0;
		 
		 n*=n-1; n/=2;
		 m*=m-1; m/=2;
		 System.out.print(n+m);
	}
}
\end{minted}
    \hfill
\begin{minted}{java}
public class Main {
	static public void main(String args[]){
		Scanner sc=new Scanner(System.in);
		 
		 int n=sc.nextInt();
		 int m=sc.nextInt();
		 if(m<2)m=0;
		 
		 n = extracted_0(n);
		 m = extracted_0(m);
		 System.out.print(n+m);
	}

	private static int extracted_0(int n) {
		n*=n-1; n/=2;
		return n;
	}
}
\end{minted}
    \caption{Original program and the expected transformation for the reuse task}
\label{refactor_example_before}
\end{figure}

Figure~\ref{refactor_example_before} shows an example program for this type of refactoring.  This example illustrates some aspects of this transformation:
\begin{enumerate}
    \item lines 9 and 10 of the original code have the same form but use different local variables, which requires the tool to understand the dataflow of those distinct variables.
    \item the extracted requires parameters to be passed the created methods, which requires the tool to consistently add the parameter to the method and arguments to each call.
\end{enumerate}

\section{Benchmark Data Characteristics}

\subsection{CodeNet} CodeNet has 1,430,135 Python programs and 74,984 Java programs.  Each of these programs was transformed as described in Section~\ref{tasks} for each transform.  Table~\ref{codenet_stats} has the total number of transformed code examples for each task.  The \texttt{Test} column shows the subset of programs in the task that we used to probe the different LLMs for code style changes.  For the programs with reuse, 906 of them had refactored methods that require passing parameters; thus the refactoring step is quite difficult. 

\begin{table}
{\small
\centering    
\begin{tabular}{|l r r|} 
 \hline
 
 Transform & \#Total  & \#Test  \\
 \hline
  List Comprehension & 179,017 & 1357 \\
  Decorator (Python) & 2,493 & 400\\
  Annotations (Java) & 750 & 75\\
  Casing (Python) & 635,306 & 1357\\
  Docstrings (Python) & 9,950 & 1357\\
 Casing (Java) & 30,887 &  250 \\
 Code Encapsulation & 1695 & 225  \\
 \hline
 \end{tabular}
    \caption{Number of transforms for each task in the CodeNet dataset.}
    \label{codenet_stats}
}
\end{table}

To test the functional correctness of the code after the transform, we ran the original code and the transformed code on the CodeNet corpus with the given inputs.  For Java, we ran on all 74,984 Java programs to ensure the transforms produce semantically correct code.

For the Python code, there were about 6,000 cases of Python programs in CodeNet where we were not able to run the code due to missing dependencies.  Another 2,722 programs failed to run on the transformed code because the library that wrote the transformed AST back to the source wrote code that caused syntax errors.  For the casing transform, 45 programs failed because the transformation involved a variable that was executed using dynamically created code (i.e., using \texttt{eval} or \texttt{exec}).  All the other transformed code for Python passed the functional tests. For Java, all changed programs executed correctly.

\subsection{BigQuery}
Although CodeNet has this desirable property of being executable, the diversity of the types of programs in CodeNet is somewhat limited.  To increase the diversity of programs useful for training LLMs we ran on Google's BigQuery dataset (bigquery-public-data.github\_repos).  As in prior work \cite{cubert}, files were deduplicated; a subset of BigQuery is the same as the \cite{cubert} dataset. Because our refactoring for Java requires compilable code, we could not take the BigQuery dataset for Java and perform transforms on it.  Therefore our Java training datasets were drawn from CodeNet.  Table~\ref{data_size} shows the characteristics of the train, validation, and test sets for BigQuery.  Note that we did not use these test sets in our evaluation, but we did create the splits so we could use training and validation for our fine-tuning.


\begin{table}
{\small
    \centering    
\begin{tabular}{|l r r r r|} 
 \hline
 Transform & \#Train & \#Val & \#Test & \#Test\\
           &         &       &        & Subset\\
 \hline
 List Comprehension & 61383 & 6821 & 477 & 477 \\
 Decorator & 146260 & 36566 & 9105 & 2001 \\
 Casing (Python) & 648540 & 162135 & 68620  & 1997\\
 Docstrings & 2281778 & 253531 & 17353 & 2001 \\ 
 \hline
 \end{tabular}
    \caption{Number of training, validation, and test examples for each task from the BigQuery datasets.}
    \label{data_size}
}
\end{table}


\section{Experiments}

\begin{table*}[ht]
{\small
\begin{center}
\begin{tabular}{llrrrrrrr}
\hline
Task & Model & CodeBLEU  & Added & Removed  & No Unexp. & No Unexp. & Passed & Passed \\
& &  & Correctly & Correctly & Removed & Added & & Correct \\
\hline
\multirow{2}{*}{List} & Starcoder & .82 & .81 & .84 & .62 & .47 & .61 &   .36     \\
\multirow{2}{*}{Comprehension} & Code Llama & .91 & .83 & .86 & .60 & .60 & .72 & .41\\
& CodeT5 & .99  & .96 & .99 & .74 & .61 & .57 & .55 \\
\hline
\multirow{2}{*}{Decorator} & Starcoder  & .66     &  .34 & - & .86 & .36 & 0.62 & .26  \\
\multirow{2}{*}{(Python)} & Code Llama & .72  & .41 & - & .92 & .44 & .55 & .23 \\
& CodeT5 & .92  & .16 & - & 0 & .13 & 0 & 0 \\

\hline
\multirow{2}{*}{Decorator} & Starcoder   &    .92    & 0 & - & .14 & 0 & .54 & 0  \\
\multirow{2}{*}{(Java)} & Code Llama & .97 &  0 & - & .50 & .02 & .73 & 0\\
& CodeT5 & .99  &  .93 & - & 1 & .88 & .46 & .41  \\

\hline
\multirow{2}{*}{Casing} & Starcoder  & .81   & .40 & .41 & .80 & .79 & .81 & .29        \\
\multirow{2}{*}{(Python)} & Code Llama & .84  & .40 & .41 & .85 & .81 & .80 & .32 \\
& CodeT5 & .96  & .48 & .54 & .63 & .25 & .31 & .12 \\
\hline

\multirow{2}{*}{Casing} & Starcoder  & .92  & .89 & .99 & .13 & .004 & .25 & 0         \\
\multirow{2}{*}{(Java)} & Code Llama & .85  & .86 & .99 & .20 & .01 & .72 & 0 \\
 & CodeT5 &  .96  & .99 & .99 & .08 & .08 & .46 & .07 \\
\hline

\multirow{2}{*}{Docstrings} & Starcoder & .67   & .97 & - &.97 &.95 &.85 & .83      \\
& Code Llama & .61  & .97 & - & .89 & .16 & .15 & .13 \\
 & CodeT5 & .67  & .98 & - & .61 & .15 & .56 & .12 \\
\hline

\multirow{2}{*}{Code encapsulation}& Starcoder & .86  & 0 & .04 & .11 & .04 & .27 & 0 \\
\multirow{2}{*}{(Java)} & Code Llama & .89 &  0 & .09 & .20 & .03 & .70 & 0 \\
 & CodeT5 &       .92  & .47 & .59 & .70 & .59 & .12 & .02 \\
\hline
\hline
\end{tabular}
\end{center}
\caption{The performance of models, showing which lines were added or removed correctly, with no unexpected additions or deletions; the table shows fractions of programs where the given model correctly computed the given metric}
\label{indi_transfer_score}
}
\end{table*}

We examined the level to which large pre-trained code models understand code well enough to change program style with few shot prompting.  To assess possible improvements in code understanding through fine-tuning, we also fine-tuned a language model but on a smaller code model due to a lack of sufficient resources to train large models.  

Large language models (LLMs) have been shown to be capable of text style transfer with in-context learning (\cite{reif-etal-2022-recipe, sun2023tsst}). To evaluate LLMs on the code style transfer task, we chose Starcoder-15.5B \cite{li2023starcoder} and codellama-34B-instruct \cite{rozière2023code} for in-context learning. Both are state-of-the-art open-source models which have been trained on multiple programming languages. We use 1-shot prompting with a carefully chosen example and a control instruction for each of the tasks. The prompts are included in the appendix. 

To examine if performance is any better from fine-tuning, we used our parallel corpora for fine-tuning a small model.  We picked CodeT5 (\cite{codet5}) for fine-tuning as it was trained using multiple code tasks. We fine-tuned CodeT5-small (60M) with two different learning rates ($2e-3$ and $2e-5$) and found that CodeT5-small with a learning rate of $2e-3$ performed the best. The max length of CodeT5 was set to 512. Each model was fine-tuned for four epochs for all tasks. 

\subsection{Model output examples}
Figure~\ref{java_casing_example} provides an example of the types of errors we observe with code LLMs.  The model (in this example both codeLLAMA) does find the variables \texttt{numofsock} and converts it correctly to camel case (which is language appropriate), and changes its declaration and uses.  However, it incorrectly changes the case of \texttt{A} and \texttt{B} as well, even though it would be unchanged in Java.  We designed a metric called DiffCorrect to capture these sorts of errors because the model could easily pass functional tests without making any of the requisite changes.  The core idea is to capture the categories of lines that are \texttt{Added Correctly} and \texttt{Removed Correctly} (green lines), but with no unexpected additions or deletions (red lines). 

\begin{figure}
\centering
\begin{minted}{java}
public class Main {
  public static void main(String[] args) {
    FastReader s = new FastReader();
    int a=s.nextInt();
    int b=s.nextInt();
    int numOfSock=1;
    int count=0;
    while(numOfSock<b)
    {
      numOfSock-=1;
      numOfSock+=a;
      count++;
    }
    System.out.println(count);
  }
}
\end{minted}
    \hfill
\begin{minted}{java}
public class Main {
  public static void main(String[] args) {
    FastReader s = new FastReader();
    int a=s.nextInt();
    int b=s.nextInt();
    int numofsock=1;
    int count=0;
    while(numofsock<b)
    {
      numofsock-=1;
      numofsock+=a;
      count++;
    }
    System.out.println(count);
  }
}
\end{minted}
    \hfill
    \begin{minted}[linenos,obeytabs,tabsize=1,escapeinside=||,fontsize=\footnotesize,  escapeinside=||]{java}
public class Main {
  public static void main(String[] args) {
    FastReader s = new FastReader();
    |\textcolor{red}{int A = s.nextInt();}|
    |\textcolor{red}{int B = s.nextInt();}|
    |\textcolor{ForestGreen}{int numOfSock = 1;}|
    int count = 0;
    |\textcolor{red}{while(numOfSock < B)}|
    {
      |\textcolor{ForestGreen}{numOfSock -= 1;}|
      |\textcolor{red}{numOfSock += A;}|
      count++;
    }
      System.out.println(count);
    }
}
    \end{minted}
    \caption{Original program, its transform, and CodeLlama output for casing changes on the same program in Java.  Highlighted lines in the original program show the lines to be changed, such that variables are camel-cased.  Note that CodeLlama correctly identified the lines to be removed and added them (in green), but also changed lines in red incorrectly.}
\label{java_casing_example}
\end{figure}

\subsection{DiffCorrect}

 As described above, we look at the changes expected for a given style transfer instance and see how well they got done while leaving code that should not be touched unchanged.  This is accomplished by the code in Figure~\ref{code-changes} which computes the DiffCorrect metric.

 \newlength\myindent
\setlength\myindent{2em}
\newcommand\bindent{%
  \begingroup
  \setlength{\itemindent}{\myindent}
  \addtolength{\algorithmicindent}{\myindent}
}
\newcommand\eindent{\endgroup}

\begin{figure}
\begin{centering}
\begin{algorithmic}
\STATE {Function {\sc CodeCompare} input, output, expected}
\STATE
\bindent
\STATE $er \gets$ {{\sc Filter}({\sc isDeletion}, {\sc Diff}(input, expected))} 
\STATE $ar \gets$ {{\sc Filter}({\sc isDeletion}, {\sc Diff}(input, output))}
\STATE $removedCorrectly \gets \left(er - ar\right) = \varnothing$
\STATE $noUnexpectedRemoved \gets \left(ar - er\right) = \varnothing$ 
\STATE
\STATE $ea \gets$ {{\sc Filter}({\sc isAddition}, {\sc Diff}(input, expected))}
\STATE $aa \gets$ {{\sc Filter}({\sc isAddition}, {\sc Diff}(input, output))}
\STATE $addedCorrectly \gets \left(ea - aa\right) = \varnothing$
\STATE $noUnexpectedAdded \gets \left(aa - ea\right) = \varnothing$ 
\eindent
\STATE
\STATE {EndFunction}
\end{algorithmic}
\end{centering}
\caption{Computing model changes}
\label{code-changes}
\end{figure}

 Figure~\ref{code-changes} has three parameters, $input$, $output$, and $expected$ which are, respectively, the original program, the model output, and the expected restyled program.  The approach is to compute diffs of diffs: $er$ gets all lines removed from $input$ in $expected$, and $ar$ gets all the lines removed from $input$ in $output$.   We then compute $\left(er - ar\right) = \varnothing$, meaning whether everything expected to be removed is indeed removed, assigning that to $removedCorrectly$.  Then $\left(ar - er\right) = \varnothing$ captures whether everything removed is expected to be removed, which is assigned to $noUnexpectedRemoved$.  There is analogous code for additions.

 Ideally, all four results are 1, meaning exactly what should be added and removed is added and removed respectively, with no unexpected changes to the code.

 We tried multiple implementations of {\tt{diff}} for this approach, and we ultimately settled on {\tt{difflib}}\footnote{\url{https://docs.python.org/3/library/difflib.html}} as providing the smallest diffs of files; a larger diff includes lines that did not in fact change, which could make our results seem worse than they are.

\subsection{Results}

Table~\ref{indi_transfer_score} shows the results of the models on the DiffCorrect metrics.  For transforms that do not involve any deletion (e.g. docstrings and decorators) removing lines is not needed so we leave that metric blank.  CodeBLEU is provided for comparison, but DiffCorrect gives us more insights into where the models go wrong.  As shown in the table, models have difficulty changing only the requisite lines of code, and leaving other lines unchanged.  Both LLMs were weaker in Java than in Python.  The most surprising result is for docstrings; powerful models such as Code Llama change the code to the point where it fails functional tests.  In the few cases we examined manually, it was because Code Llama produced docstrings that would not parse as a string literal.

Fine-tuning did help a much smaller model (Code T5) perform list comprehension better, and for casing, it helped identify the lines to be changed correctly. However, fine-tuning did not always help with the tendency to edit incorrect lines. We also see how the fine-grained nature of the DiffCorrect metric helps us understand exactly where models go wrong compared to traditional metrics such as CodeBLEU.  CodeBLEU combined with pass rates would have provided a misleading picture of the ability of these models to transform code.

\section{Conclusions and Future Work}
We evaluated the capability of current code LLMs to perform code style changes because this is an important requirement for many organizations.  To our surprise, existing code LLMs could not perform this task well.  To help LLMs perform better on such a task, we built benchmarks and defined a new metric called DiffCorrect that can help in building better models.  Strengthening the ability of LLMs to perform these types of transformations is a target for future work. 
\clearpage

\bibliographystyle{named}
\bibliography{ijcai24}

\end{document}